\documentclass[aps,prl,twocolumn,superscriptaddress,10pt,showpacs]{revtex4}

\usepackage{amsmath}
\usepackage{graphicx}
\usepackage{bm}

\begin{document}
\title{First Measurement of the Neutron $\bm{\beta}$-Asymmetry with Ultracold Neutrons}

\author{R.~W.~Pattie Jr.}
\affiliation{Department of Physics, North Carolina State University, Raleigh, North Carolina 27695, USA}
\affiliation{Triangle Universities Nuclear Laboratory, Durham, North Carolina 27708, USA}
\author{J.~Anaya}
\affiliation{Los Alamos National Laboratory, Los Alamos, New Mexico 87545, USA}
\author{H.~O.~Back}
\affiliation{Department of Physics, North Carolina State University, Raleigh, North Carolina 27695, USA}
\affiliation{Triangle Universities Nuclear Laboratory, Durham, North Carolina 27708, USA}
\author{J.~G.~Boissevain}
\affiliation{Los Alamos National Laboratory, Los Alamos, New Mexico 87545, USA}
\author{T.~J.~Bowles}
\affiliation{Los Alamos National Laboratory, Los Alamos, New Mexico 87545, USA}
\author{L.~J.~Broussard}
\affiliation{Triangle Universities Nuclear Laboratory, Durham, North Carolina 27708, USA}
\affiliation{Department of Physics, Duke University, Durham, North Carolina 27708, USA}
\author{R.~Carr}
\affiliation{Kellogg Radiation Laboratory, California Institute of Technology, Pasedena, California 91125, USA}
\author{D.~J.~Clark}
\affiliation{Los Alamos National Laboratory, Los Alamos, New Mexico 87545, USA}
\author{S.~Currie}
\affiliation{Los Alamos National Laboratory, Los Alamos, New Mexico 87545, USA}
\author{S.~Du}
\affiliation{Department of Physics, North Carolina State University, Raleigh, North Carolina 27695, USA}
\author{B.~W.~Filippone}
\affiliation{Kellogg Radiation Laboratory, California Institute of Technology, Pasedena, California 91125, USA}
\author{P.~Geltenbort}
\affiliation{Institut Laue-Langevin, 38042 Grenoble Cedex 9, France}
\author{A.~Garc\'{\i}a}
\affiliation{Physics Department, University of Washington, Seattle, Washington 98195, USA}
\author{A.~Hawari}
\affiliation{Department of Physics, North Carolina State University, Raleigh, North Carolina 27695, USA}
\author{K.~P.~Hickerson}
\affiliation{Kellogg Radiation Laboratory, California Institute of Technology, Pasedena, California 91125, USA}
\author{R.~Hill}
\affiliation{Los Alamos National Laboratory, Los Alamos, New Mexico 87545, USA}
\author{M.~Hino}
\affiliation{Research Reactor Institute, Kyoto University, Kumatori, Osaka, 590-0401, Japan}
\author{S.~A.~Hoedl}
\affiliation{Physics Department, University of Washington, Seattle, Washington 98195, USA}
\affiliation{Physics Department, Princeton University, Princeton, New Jersey 08544, USA}
\author{G.~E.~Hogan}
\affiliation{Los Alamos National Laboratory, Los Alamos, New Mexico 87545, USA} 
\author{A.~T.~Holley}
\affiliation{Department of Physics, North Carolina State University, Raleigh, North Carolina 27695, USA}
\author{T.~M.~Ito}
\affiliation{Los Alamos National Laboratory, Los Alamos, New Mexico 87545, USA}
\affiliation{Kellogg Radiation Laboratory, California Institute of Technology, Pasedena, California 91125, USA}
\author{T.~Kawai}
\affiliation{Research Reactor Institute, Kyoto University, Kumatori, Osaka, 590-0401, Japan}
\author{K.~Kirch}
\affiliation{Los Alamos National Laboratory, Los Alamos, New Mexico 87545, USA}
\author{S.~Kitagaki}
\affiliation{Tohoku University, Sendai 980-8578, Japan}
\author{S.~K.~Lamoreaux}
\affiliation{Los Alamos National Laboratory, Los Alamos, New Mexico 87545, USA}
\author{C.-Y.~Liu}
\affiliation{Physics Department, Princeton University, Princeton, New Jersey 08544, USA}
\author{J.~Liu}
\affiliation{Kellogg Radiation Laboratory, California Institute of Technology, Pasedena, California 91125, USA}
\author{M.~Makela}
\affiliation{Los Alamos National Laboratory, Los Alamos, New Mexico 87545, USA}
\affiliation{Department of Physics, Virginia Tech, Blacksburg, Virginia 24061, USA}
\author{R.~R.~Mammei}
\affiliation{Department of Physics, Virginia Tech, Blacksburg, Virginia 24061, USA}
\author{J.~W.~Martin}
\affiliation{Kellogg Radiation Laboratory, California Institute of Technology, Pasedena, California 91125, USA}
\affiliation{Department of Physics, University of Winnipeg, Winnipeg, MB R3B 2E9, Canada}
\author{D.~Melconian}
\affiliation{Physics Department, University of Washington, Seattle, Washington 98195, USA}
\affiliation{Cyclotron Institute, Texas A$\&$M University, College Station, Texas 77843, USA}
\author{N.~Meier}
\affiliation{Department of Physics, North Carolina State University, Raleigh, North Carolina 27695, USA}
\author{M.~P.~Mendenhall}
\affiliation{Kellogg Radiation Laboratory, California Institute of Technology, Pasedena, California 91125, USA}
\author{C.~L.~Morris}
\affiliation{Los Alamos National Laboratory, Los Alamos, New Mexico 87545, USA}
\author{R.~Mortensen}
\affiliation{Los Alamos National Laboratory, Los Alamos, New Mexico 87545, USA}
\author{A.~Pichlmaier}
\affiliation{Los Alamos National Laboratory, Los Alamos, New Mexico 87545, USA}
\author{M.~L.~Pitt}
\affiliation{Department of Physics, Virginia Tech, Blacksburg, Virginia 24061, USA}
\author{B.~Plaster}
\affiliation{Kellogg Radiation Laboratory, California Institute of Technology, Pasedena, California 91125, USA}
\affiliation{Department of Physics and Astronomy, University of Kentucky, Lexington, Kentucky 40506, USA}
\author{J.~C.~Ramsey}
\affiliation{Los Alamos National Laboratory, Los Alamos, New Mexico 87545, USA}
\author{R.~Rios}
\affiliation{Los Alamos National Laboratory, Los Alamos, New Mexico 87545, USA}	
\affiliation{Department of Physics, Idaho State University, Pocatello, Idaho 83209, USA}
\author{K.~Sabourov}
\affiliation{Department of Physics, North Carolina State University, Raleigh, North Carolina 27695, USA}
\author{A.~Sallaska}
\affiliation{Physics Department, University of Washington, Seattle, Washington 98195, USA}
\author{A.~Saunders}
\affiliation{Los Alamos National Laboratory, Los Alamos, New Mexico 87545, USA}
\author{R.~Schmid}
\affiliation{Kellogg Radiation Laboratory, California Institute of Technology, Pasedena, California 91125, USA}
\author{S.~Seestrom}
\affiliation{Los Alamos National Laboratory, Los Alamos, New Mexico 87545, USA}
\author{C.~Servicky}
\affiliation{Department of Physics, North Carolina State University, Raleigh, North Carolina 27695, USA}
\author{S.~K.~L.~Sjue}
\affiliation{Physics Department, University of Washington, Seattle, Washington 98195, USA}
\author{D.~Smith}
\affiliation{Department of Physics, North Carolina State University, Raleigh, North Carolina 27695, USA}
\author{W.~E.~Sondheim}
\affiliation{Los Alamos National Laboratory, Los Alamos, New Mexico 87545, USA}
\author{E.~Tatar}
\affiliation{Department of Physics, Idaho State University, Pocatello, Idaho 83209, USA}	
\author{W.~Teasdale}
\affiliation{Los Alamos National Laboratory, Los Alamos, New Mexico 87545, USA}
\author{C.~Terai}
\affiliation{Department of Physics, North Carolina State University, Raleigh, North Carolina 27695, USA}
\author{B.~Tipton}
\affiliation{Kellogg Radiation Laboratory, California Institute of Technology, Pasedena, California 91125, USA}
\author{M.~Utsuro}
\affiliation{Research Reactor Institute, Kyoto University, Kumatori, Osaka, 590-0401, Japan}
\author{R.~B.~Vogelaar}
\affiliation{Department of Physics, Virginia Tech, Blacksburg, Virginia 24061, USA}
\author{B.~W.~Wehring}
\affiliation{Department of Physics, North Carolina State University, Raleigh, North Carolina 27695, USA}
\author{Y.~P.~Xu}
\affiliation{Department of Physics, North Carolina State University, Raleigh, North Carolina 27695, USA}
\author{A.~R.~Young}
\affiliation{Department of Physics, North Carolina State University, Raleigh, North Carolina 27695, USA}
\affiliation{Triangle Universities Nuclear Laboratory, Durham, North Carolina 27708, USA}
\author{J.~Yuan}
\affiliation{Kellogg Radiation Laboratory, California Institute of Technology, Pasedena, California 91125, USA}

\collaboration{The UCNA Collaboration}

\date{\today}

\begin{abstract}
We report the first measurement of angular correlation parameters in
neutron $\beta$-decay using polarized ultracold neutrons (UCN).  We
utilize UCN with energies below about 200 neV, which we guide and
store for $\sim 30$ s in a Cu decay volume.  The $\vec{\mu}_n \cdot
\vec{B}$ potential of a static 7 T field external to the decay volume
provides a 420 neV potential energy barrier to the spin state parallel
to the field, polarizing the UCN before they pass through an
adiabatic fast passage (AFP) spin-flipper and enter a decay volume,
situated within a 1 T, $2\times2\pi$ superconducting solenoidal
spectrometer.  We determine a value for the $\beta$-asymmetry
parameter $A_0$, proportional to the angular correlation between the
neutron polarization and the electron momentum, of $A_0 = -0.1138 \pm
0.0051$.
\end{abstract}


\maketitle 

Measurements of neutron $\beta$-decay observables provide fundamental
information on the parameters characterizing the weak interaction of
the nucleon.  Results from such measurements can be used to extract a
value for the CKM quark-mixing matrix element $V_{ud}$ and impact
predictions for the solar neutrino flux, big bang nucleosynthesis, the
spin content of the nucleon, and tests of the Goldberger-Treiman
relation \cite{Czarn04}.  High-precision results also place
constraints on various extensions to the standard model, such as
supersymmetry \cite{Musol02} and left-right symmetries \cite{Sever06}.
Angular correlation measurements in neutron $\beta$-decay have been
performed with thermal and cold neutron beams~\cite{Nico06}, including
all previously reported measurements of the $\beta$-asymmetry
\cite{Abele02, Liaud97, Yerol91, Bopp86, Krohn75}. The use of UCN for
these measurements provides a different and powerful approach to
controlling key sources of systematic errors in measurements of
polarized neutron $\beta$-decay: the preparation of highly polarized
neutrons and the backgrounds intrinsic to the neutron $\beta$-decay
sample.\\
\indent The $\beta$-asymmetry results from the angular correlation
between the neutron spin and the electron momentum, where
the angular distribution of emitted electrons is
\begin{equation}
W(E_e,\theta) = F(E_e) ( 1+A\langle P\rangle \beta \cos\theta).
\label{eq:angular}
\end{equation}
Here, $E_e$ and $\beta$ are the electron energy and velocity
relative to $c$, $F(E_e)$ is the allowed shape of the electron energy
spectrum, $\langle P \rangle$ is the neutron polarization, and $\theta$ is
the angle between the neutron spin and the electron momentum.  The
magnitude of the asymmetry is the $A$ parameter
\cite{Hols74,Gardne01,Jack57}, given by $A = A_{0}(1 + a_0 +
a_{-1}/E_e + a_{+1} E_e)(1+\delta)$, where $A_{0} =
-2\lambda(\lambda+1)/(1+3\lambda^2)$.  The
recoil order terms $a_0$, $a_{-1}$, $a_{+1}$ and the radiative correction
$\delta$ are specified in the standard model in
terms of six effective coupling constants, with leading-order
contributions from two form factors ($g_v$ and $g_a$, the vector and
axial-vector coupling constants with $\lambda \equiv g_a / g_v$) and one form factor which only
appears in the recoil order terms
($g_{wm}$, the weak magnetism coupling constant).  The other three
form factors are expected to be negligible at the level of precision
of this work. Very precise estimates for $g_{v}$ and $g_{wm}$ are
available in the standard model, but precise theoretical predictions
for $g_a$ are not available. Other than $g_a$, limiting
theoretical uncertainties are below the $0.1\%$ level, and stem from
hadronic loop contributions to radiative corrections~\cite{Marci06}.

\begin{figure}[tbp] 
  \centering
  \includegraphics[angle=270,scale=0.34,clip=]{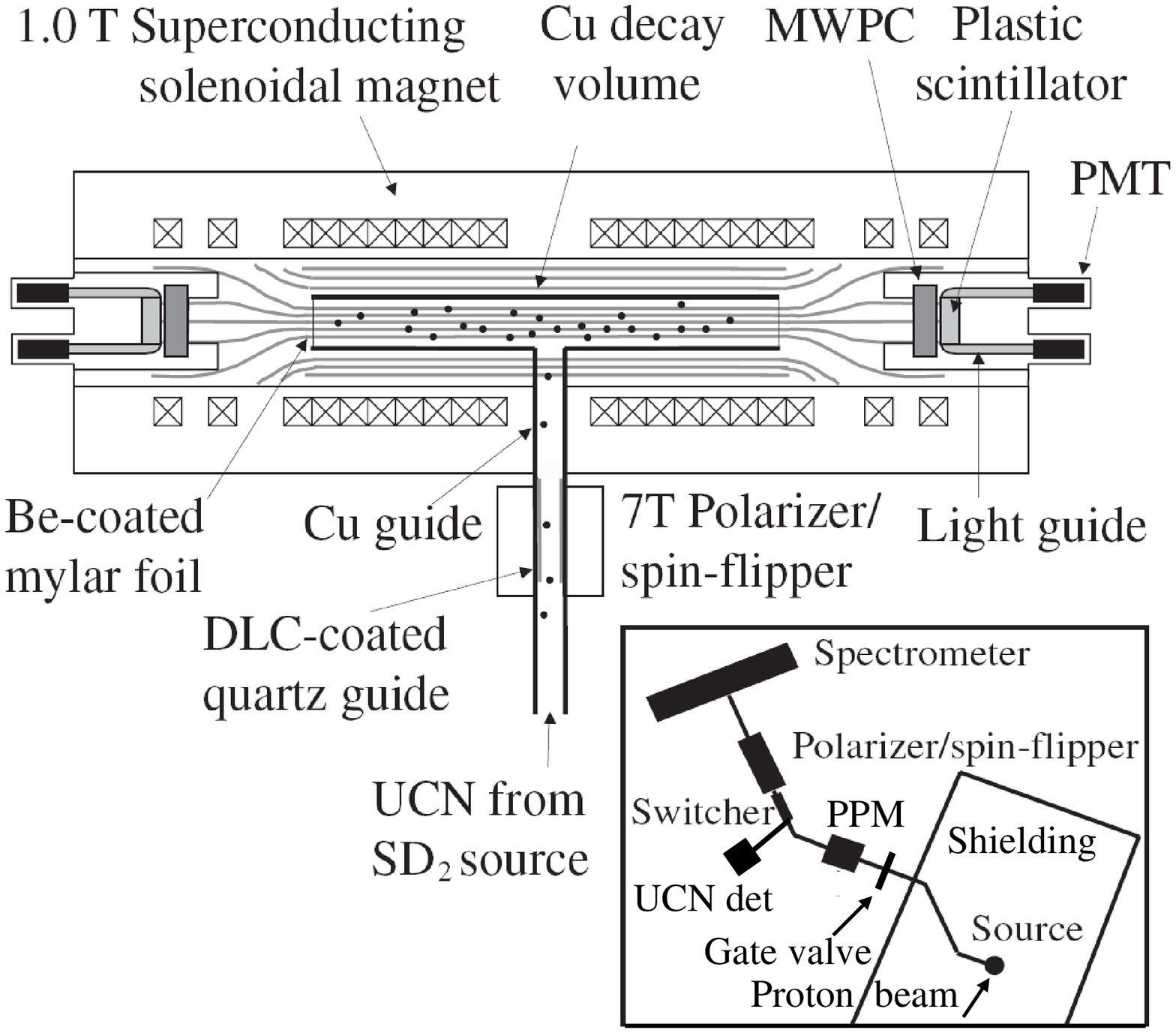}
  \caption{Schematic (not to scale) of the $\beta$-asymmetry experiment.
  Inset depicts the layout of the UCN source and transport guides.}
  \label{fig:ucnas}
\end{figure}
\indent UCN are defined as neutrons with energies low enough ($\alt$
340 neV) that they
undergo total external reflection from an effective
potential barrier $E_{\mathrm{Fermi}}$ at some material surfaces
\cite{Golub} and can therefore be stored in material bottles.
We produced UCN in a solid deuterium (SD$_2$) source \cite{Saunders00}
closely coupled to a tungsten spallation target in the 800 MeV proton
beam at the Los Alamos Neutron Science Center (LANSCE).  Protons were
delivered in 28 $\mu$C pulses, once every 17 s, with the spallation
neutrons moderated in cold ($\sim 20$~K) polyethylene surrounding the
source.  UCN were created via downscattering of the resulting cold
neutron flux in 5~K SD$_2$.  After each proton pulse, the emerging UCN
passed through a valve located above the SD$_2$.  This valve was then
immediately closed, loading a storage volume (20~l vertical volume and
40~l of guides) above the SD$_2$ with UCN, at an estimated density of
10 UCN cm$^{-3}$ at the exit of the spallation source shielding.  The
vertical storage volume was coupled by about 5 m of electropolished
stainless steel guides (with two $45^\circ$ bends to limit backgrounds)
to a gate valve just beyond the shielding, and then ultimately to a
switcher which allowed the guides comprising the $\beta$-asymmetry
experiment (schematic shown in Fig.\ \ref{fig:ucnas}) to be connected
either to the UCN source or a $^3$He UCN detector for depolarization
measurements, described below.\\
\indent One of the primary advantages of the low UCN energy is the
ability to highly polarize a UCN population using the $\vec{\mu}_n
\cdot \vec{B}$ potential associated with a static magnetic field,
which amounts to $\pm60$ neV T$^{-1}$ for neutron spins
aligned($+$)/anti-aligned($-$) with the field. The UCN flux
was polarized via this effect, using a longitudinal
6~T field in a pre-polarizer magnet (PPM) and a 7 T field in the
primary polarizer/spin-flipper magnet.
The polarization of the subsequent
population was maintained by polarization-preserving
electropolished Cu ($E_{\mathrm{Fermi}} \approx 168$ neV) guides and a
100-cm long diamond-like carbon (DLC) coated quartz section
($E_{\mathrm{Fermi}}>200$~neV) passing through the center of a
resonant ``birdcage'' rf cavity 
used for
adiabatic fast passage (AFP) spin flipping of the UCN.
Holding fields were established by the homogeneous 1~T
field required for operation of the spin-flipper, the 1~T field in
the spectrometer, and the superposition of the magnets' fringe fields
in the connecting region. \\
\indent Prior to the $\beta$-asymmetry run, a measurement of the AFP
spin-flipper efficiency was made by placing the PPM magnet in
the section which normally connects the polarizer/spin-flipper magnet
with the spectrometer, creating a crossed polarizer analyzer with the
PPM utilized as the analyzer. In order to increase
the analyzing power of the PPM (whose maximum field was limited to 6
T), a $\sim13\;\mu$m thick Cu-coated kapton foil was placed in the
high-field region. The transmission through the system
was measured to be $0.40(5)\%$, placing a
lower limit of $99.1\%$ on the initial UCN polarization and a lower
limit of $99.6\%$ on the spin-flipper efficiency.
Based on spectral measurements and limitations of the PPM/foil
analyzer we expect the actual values of the initial UCN polarization
and spin-flipper efficiency to be higher; however, the lower limits
presented here were used in our estimates of polarization-related
systematics. \\
\indent In order to measure the depolarization on a run-by-run basis,
the measurement cycle consisted of a background run (with proton beam
on, but gate valve closed, resulting in no UCN in the spectrometer,
$\sim 720$~s in duration) followed by the $\beta$-asymmetry
measurement for some spin-state ($\sim 3600$~s), and then a
measurement of the depolarized population ($\sim 240$~s).  The
depolarization measurement consisted of first closing the gate valve
while simultaneously connecting the guides at the upstream side of the
experiment to the $^3$He UCN detector using the switcher. This
cleaning phase, lasting 150~s, allowed the number of correctly
polarized UCN downstream of the 7 T polarizing field to be measured.
Next, the depolarized population present in the experiment was
unloaded and counted by changing the state of the spin-flipper.
Counting during this unloading cycle was carried out for 100~s after
which the next measurement cycle was started.  The depolarized
contamination at the end of each individual run was consistent with
zero.  By combining all available runs, correcting for depolarized UCN
detection efficiency, and attributing the crossed polarized analyzer
result entirely to unpolarized UCN, we were able to place an upper
limit (1$\sigma$) of 0.65\% on the depolarized fraction present during
any individual run.  We expect that better characterization of our
system combined with significant improvements in background and
statistics during depolarized UCN counting should result in a limit
below 0.2\% in the future. \\
\indent After transport to the spectrometer, the UCN were confined to
a 300-cm long, 12-cm diameter Cu decay trap tube with end-cap foils
consisting of 2.5-$\mu$m thick mylar foils coated with 300 nm of
Be. Storage times of $\sim 30$ s were achieved.  A plastic collimator
with an inner diameter of 11.7 cm, which also functioned as the
end-cap foil mount, suppressed contamination from electrons scattering
from the Cu decay trap or resulting from neutron capture on the Cu
walls.  The spectrometer's 1~T solenoidal field was oriented along the
decay trap axis and provided for $2 \times 2\pi$ collection of the
$\beta$-decay electrons, which spiraled along the field lines toward
one of two identical electron detector packages \cite{fields}.  This
expansion of the field from 1~T in the decay region to 0.6~T in the
detector region
suppresses electron backscattering (scattering at angles $>90^{\circ}$
relative to incidence) via two mechanisms.  First, pitch angles of
$90^{\circ}$ in the 1 T region map to $51^{\circ}$ in the 0.6 T
region.  Second, electrons which backscatter at
pitch angles $\theta > 51^{\circ}$ are
reflected via the magnetic mirror effect back into the same detector. \\
\indent The two electron detector packages
\cite{Ito07,Plaster08} each consisted of a low-pressure (132 mbar)
multiwire proportional chamber (MWPC) backed by a 3.5 mm thick, 15 cm
diameter plastic scintillator, and were spaced 4.4~m apart.
Requiring a coincidence between the same-side MWPC and scintillator
led to a significant reduction in backgrounds.  The MWPCs
also permitted reconstruction (with 2 mm resolution) of the
events' transverse coordinates, for the rejection of events
originating near the edge of the collimator.  The position sensitivity
of our MWPCs ensured ``edge effects'' associated with electron scatter from
the walls of our decay trap or collimator edges were negligible.
The MWPCs were
separated from the spectrometer vacuum and the scintillator volumes by
identical 25~$\mu$m thick mylar
entrance and exit windows. \\
\indent The scintillator generated the trigger and provided for the
energy measurement.  Energy calibrations were performed \textit{in
situ} every 6--8 hours, using a $^{113}$Sn source of
conversion electrons which could be inserted and retracted
from the spectrometer's fiducial volume.  Gain variations between
calibrations were monitored and corrected for by comparing the
response of the scintillator and an external photomultiplier tube
(PMT) to a pulsed LED source.  The response of the external PMT to the
LED was normalized to the spectrum of gamma rays from a $^{60}$Co
source in a NaI crystal coupled to this PMT.

The linearity of the detector response was studied at the conclusion
of the run with a limited selection of conversion-electron sources
($^{113}$Sn and $^{207}$Bi), effectively calibrating the detectors.
Analysis of the implementation of our calibration via Monte
Carlo suggests a (conservative) 1.5\% uncertainty in the value for
$A_{0}$ due to possible errors in our electron energy reconstruction.
Note that with these calibration studies, the
resolution of the scintillator was shown to be 5.6\% at 1~MeV.

\indent The spectrometer was surrounded by a cosmic-ray muon veto
system consisting of proportional gas tubes and plastic scintillator.
Proton beam-related backgrounds were suppressed with timing cuts.  A
significant potential advantage of UCN over cold neutron beams for
angular correlation measurements stems from the relatively small
number of neutrons present in the decay geometry and guides.  Because
the absolute efficiency of an MWPC-scintillator coincidence detection
of neutron-generated gamma-ray backgrounds within the fiducial volume
is extremely small compared to the nearly 100\%
electron-detection efficiency, UCN-generated backgrounds were
negligible.\\
\begin{table}[t!]
\caption{Fractional systematic corrections to $A_{0}$.}
 \begin{ruledtabular}
  \begin{tabular}{lcc}
     Systematic & Correction & Uncertainty \\
     \hline
     Polarization & ---& 0.013 \\
     UCN-Induced Backgrounds & --- & 0.002 \\ \hline
     Electron Detector Effects &  &\\
     ~~Response/Linearity & ---  & 0.015 \\
     ~~Width/Pedestal & ---  & 0.001 \\
     ~~Gain Drifts & ---  & 0.002 \\ \hline
     Electron Trajectories & & \\
     ~~Angle Effects & $-0.016$ & 0.005\\
     ~~Backscattering & 0.011 & 0.004\\ \hline
     Total        & $-5 \times 10^{-3}$& 0.021 \\
  \end{tabular}
\label{tab:corrs}
 \end{ruledtabular}
\end{table}
\indent Misidentification of backscattering events is one of the
largest contributions to the total fractional systematic correction to
the value for $A_{0}$ reported here (see
Table~\ref{tab:corrs}). Electrons triggering both scintillators within
a 100~ns acceptance window comprised 2.5(3)$\%$ of all events, with
the initial direction of incidence determined by the scintillators'
relative trigger times.  The bias to the asymmetry from events with
transit times greater than 100 ns is small, $< 10^{-5}$.  Another
class of backscattering events, comprising 1.4(2)\% of the event
fraction, were those events which triggered only one of the
scintillators, but deposited energy in both MWPCs.  Comparison of the
energy deposition in the MWPCs determines the initial direction of
incidence for these events, with the efficiency for this
identification calculated in Monte Carlo to be $\sim 80$\%.  Finally,
events which backscattered from either the decay trap end-cap foils or
the MWPC entrance windows could not be identified in data analysis,
and are termed ``missed'' backscattering events.  Corrections for
these missed backscattering events were separately performed using the
GEANT4 \cite{GEANT4} and PENELOPE \cite{PENELOPE} simulation packages,
and found to be $1.1(4)$\%.  Previous studies of these Monte Carlo
backscattering calculations suggest a (conservative) $30\%$
uncertainty in the missed backscattering correction \cite{JMartin}.\\
\begin{figure}[t!]
  \centering
        \includegraphics[scale=0.40,angle=270]{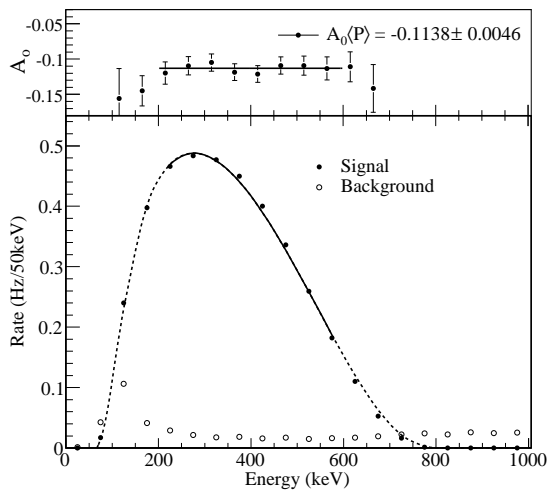}
  \caption{Top panel: Values for $A_{0}$ extracted from the best fit,
           varying the parameter $\lambda$, plotted as a function of the
           reconstructed $\beta$-decay energy (i.e., after accounting
           for energy loss; see text for details).  The value of $\delta$
           is below $0.1\%$ \cite{Gluck} and therefore negligible for this
           study.  Bottom panel:
           Reconstructed $\beta$-decay energy spectrum, summed over
           detectors, compared with the
           expectation from Monte Carlo (dashed line).  Solid line
           indicates analysis region.}
  \label{fig:p_spec}
\end{figure}
Linear parametrizations of energy loss in the decay trap end-cap foils
and the MWPC entrance/exit windows and interior were employed to
reconstruct, on an event-by-event basis for each event type, the
$\beta$-decay energy from the measured energy deposition in the
scintillator(s).  The resulting energy
spectrum, compared to background, is shown in Fig.\ \ref{fig:p_spec}.
Disagreement between the shape of the reconstructed energy spectrum
and that expected from simulation is $<3\%$ over the chosen analysis
region of 200--600 keV.  The lower limit of 200 keV was constrained by
energy loss for backscattering events triggering both
scintillators. The upper limit of 600 keV was an optimization of the
signal-to-noise ratio, seen in Fig.~\ref{fig:p_spec} to be 21:1 from
200--600 keV, over 0.17 Hz of background.  Note that the
background-subtracted rate above the $\beta$-decay endpoint was
consistent with zero. \\
\indent Due to the angle dependence of the energy loss, the average
value of $\cos\theta$ deviates from $1/2$ and varies as a function of
energy.  The correction to the asymmetry for such ``angle effects''
was calculated in Monte Carlo to be $-1.6(5)\%$.  Effects due to
differences in loading efficiencies for the two spin states and
detector efficiencies cancel in a super ratio of rates, $S(E_e) =
r(E_e)_1^{\uparrow}r(E_e)_2^{\downarrow}/r(E_e)_1^{\downarrow}
r(E_e)_2^{\uparrow}$,
where $r(E_e)_{1(2)}^{\uparrow(\downarrow)}$ was the rate measured in
detector 1(2) when the spin-flipper was on(off).  The experimental
asymmetry is then $A(E_e) =
(1-\sqrt{S(E_e)})/(1+\sqrt{S(E_e)})$.  After correcting for
backscattering and angle effects, $A_{0}$ was extracted from a
one-parameter fit.  With these corrections we find
$A_{0} = -0.1138(46)(21)$, where the first
(second) uncertainty is statistical (systematic).
Results for $A_{0}$ from independent analyses utilizing
PENELOPE or GEANT4 for calculation of the corrections for energy
loss and backscattering agreed to 0.26\%. \\
\indent
In summary, we have demonstrated the first-ever measurement of
a neutron $\beta$-decay angular correlation parameter with UCN.
Subsequent improvements in the UCN transport to the spectrometer will
lead to a reduction in the statistical error.  Significant
improvements in the limits placed on the fraction of depolarized UCN
should follow from improved statistics for depolarization studies and
improved operating parameters for our $^{3}$He UCN detectors.  A
transition to thinner decay trap end-cap foils and MWPC windows will
reduce the magnitude of the systematic corrections for backscattering
and angle effects, and a greater selection of calibration sources will
be employed for more extensive studies of the detector response. \\
\indent This work was supported in part by the Department of Energy,
National Science Foundation, and Los Alamos National Laboratory LDRD.
We acknowledge helpful discussions with A.\ P.\ Serebrov.


\begin{thebibliography}{999}
\bibitem{fields} In the current geometry, UCN are confined to the
decay trap, located in a uniform field region, ensuring magnetic
mirroring corrections to $A_0$ due to UCN decays in non-uniform fields
are below the $0.1\%$ level.
\end{thebibliography}
\end{document}